\documentclass[sigconf]{acmart}

\renewcommand\footnotetextcopyrightpermission[1]{} 
\usepackage{algorithmic}
\usepackage{graphicx}
\usepackage{textcomp}
\usepackage{xcolor}
\def\BibTeX{{\rm B\kern-.05em{\sc i\kern-.025em b}\kern-.08em
    T\kern-.1667em\lower.7ex\hbox{E}\kern-.125emX}}
\usepackage{listings}
\usepackage{stfloats}
\usepackage{subcaption}

\lstnewenvironment{mycode}[1][]{
  \lstset{#1}
  \setcounter{lstlisting}{\value{figure}}
}{\addtocounter{figure}{1}}

\lstnewenvironment{mycode-s}[1][]{
  \lstset{#1}
}{}

\usepackage{xcolor}

\begin{document}

\title{Clueless: A Tool \underline{C}haracterising Va\underline{lue}s \underline{L}eaking as Addr\underline{ess}es}

\author{Xiaoyue Chen}
\affiliation{%
  \institution{Uppsala University}
  \city{Uppsala}
  \country{Sweden}
}
\email{xiaoyue.chen@it.uu.se}

\author{Pavlos Aimoniotis}
\affiliation{%
  \institution{Uppsala University}
  \city{Uppsala}
  \country{Sweden}
}
\email{pavlos.aimoniotis@it.uu.se}

\author{Stefanos Kaxiras}
\affiliation{%
  \institution{Uppsala University}
  \city{Uppsala}
  \country{Sweden}
}
\email{stefanos.kaxiras@it.uu.se}

\begin{abstract}

  Clueless is a binary instrumentation tool that characterises
  explicit cache side channel vulnerabilities of programs. It detects
  the transformation of data values into addresses by tracking dynamic
  instruction dependencies. Clueless tags data values in memory if it
  discovers that they are used in address calculations to further
  access other data.

  Clueless can report on the amount of data that are used as addresses
  at each point during execution. It can also be specifically
  instructed to track certain data in memory (e.g., a password) to see
  if they are turned into addresses at any point during execution. It
  returns a trace on how the tracked data are turned into addresses,
  if they do.

  We demonstrate Clueless on SPEC 2006 and characterise, for the first
  time, the amount of data values that are turned into addresses in
  these programs. We further demonstrate Clueless on a micro benchmark
  and on a case study. The case study is the different implementations
  of AES in OpenSSL: T-table, Vector Permutation AES (VPAES), and
  Intel Advanced Encryption Standard New Instructions (AES-NI).
  Clueless shows how the encryption key is transformed into addresses
  in the T-table implementation, while explicit cache side channel
  vulnerabilities are note detected in the other implementations.

\end{abstract}



\maketitle
\pagestyle{plain}

\section{Introduction}

Cache side-channel attacks leak information through a
microarchitectural covert channel -- the cache. By observing changes in
the shared cache state, a spy process can bypass process isolation and
read secret data from a victim process. Cache side-channel attacks
have been demonstrated on processors of different architectures and on
different algorithms, e.g., RSA~\cite{yarom2014flush+},
AES~\cite{aciiccmez2007cache, osvik2006cache, bonneau2006cache,
  gullasch2011cache}, and ElGamal~\cite{liu2015lastlevelcache}.
\emph{Speculative} side-channel attacks such as
Spectre~\cite{spectre:SP2019}, Meltdown~\cite{lipp2018meltdown} and
their
variants~\cite{smotherspectre:CCS2019,Kiriansky:ARXIV2018,storebypass:WWW2018,koruyeh2018spectre,maisuradze2018ret2spec,Aimoniotis+:CAL2021-fsi}
have caused major changes on how the architecture community view
security. These attacks exploit speculative instructions that are to
be squashed (e.g., instructions in mispredicted branches) to access
and then transmit secret data over the shared cache. While
non-speculative cache side-channel attacks could usually be mitigated
by improving the implementations of vulnerable algorithms (e.g., avoid
using secret data to look up in large tables), the speculative
variants of them are difficult to prevent by changing software
implementations because the information leakage happens in
speculation.

Fig.~\ref{lst:spectre} shows Spectre Variant 1 where an attacker can
exploit the branch misprediction to access arbitrary program data and
transmit the secret over a shared cache~\cite{spectre:SP2019}. The
victim program is correctly implemented with the appropriate bound
check, yet it is still vulnerable due to speculative execution.

\begin{mycode}[
  caption={Spectre Variant 1.},
  label={lst:spectre},
  float=bp
  ]
uint8 A[10];
uint8 B[256*64];
void victim (size_t addr) {
  if (addr < 10) {  // mispredicted branch
    uint8 val = A[addr];  // secret accessed
    ... = B[64*val];  // secret transmitted
  }
}
\end{mycode}

Speculative side-channel attacks have found to be an enormous security
threat. Different hardware approaches have been proposed to protect
against them. For example, InvisiSpec~\cite{invisispec:MICRO2018} and
GhostMinion~\cite{ghostminion:MICRO2021} makes speculation invisible
in the data cache hierarchy using additional speculative buffers so
that secrets cannot be transmitted over cache channels. Delay-on-Miss
(DoM)~\cite{dom:ISCA2019} delays all speculative loads that miss in
data cache and thus prevent the observable timing differences, while
Speculative Taint Tracking (STT)~\cite{stt:MICRO2019} 
focuses on blocking only the transmitter instruction.
STT uses dynamic information flow
tracking (DIFT) to taint secret data. It allows to forward the results
of speculative instructions if they cannot leak secrets via any
potential covert channels.

This work does not propose new mitigation methods for speculative
side-channel attacks. Instead, we intend to understand how prevalent
these vulnerabilities are in programs from a new perspective.
Side-channel attacks rely on a fundamental programming feature to leak
the value of secrets -- \emph{the transformation of data values into memory
addresses}. Besides the victim program in Fig.~\ref{lst:spectre},
consider for example sorting, hashing, or many other algorithms that
create addresses based on data values. While we understand the
mechanism that leaks data as addresses, there is no clear
indication of how serious the problem is in our workloads: How many
values do ``leak'' as addresses in a given application?

This work aims to shed some light on how exposed are we to the
potential vulnerability. \emph{Clueless} is a tool (based on binary
re-writing) that tracks dynamic instruction dependencies and tags data
values in memory if it discovers that they are used in address
calculations to further access other data.

Clueless can be used in two modes: \emph{aggregating mode}, where it
reports on the amount of data that are used as addresses at each point
during execution, and \emph{tracking mode} where the tool is
specifically asked to track certain data in memory (e.g., a password)
to see if they are turned into addresses at any point during
execution. Tracking mode returns a trace on how the tracked data are
turned into addresses, if they do.

We demonstrate Clueless in aggregating mode on SPEC 2006 and
characterise, for the first time, the amount of data values that are
turned into addresses in these programs. We further demonstrate
Clueless in tracking mode on a micro benchmark and on a case study.
The case study is the different implementations of AES in OpenSSL:
T-table, Vector Permutation AES (VPAES), and Intel Advanced Encryption
Standard New Instructions (AES-NI). The T-table AES implementation can
be easily broken with a cache side-channel attack (e.g., Prime+Probe),
but VPAES and AES-NI are immune to cache-timing attacks. Clueless
readily shows how the encryption key is transformed into addresses in
the T-table implementation and a lack of the corresponding
transformations in the other two implementations.

\section{Methods}


Clueless is a dynamic instrumentation tool that analyses instructions
at run-time to track \emph{values} that leak as memory addresses.
Values are data that should not be used, directly or indirectly, as
memory addresses, e.g., password hashes, private encryption keys. A
value can leak as a memory address when there is information flow from
the value to a memory address. The scope of the tool is limited to
detecting data-flow: it tracks data dependences but disregards control
dependences. In other words, Clueless is able to detect explicit
channels~\cite{stt:MICRO2019}, a value that is used as an address on a
load instruction, but not implicit channels~\cite{stt:MICRO2019},
where the value is leaked through control flow interaction.
Assume \lstinline{secret} is a value, the leakage in the code in
Fig.~\ref{lst:ctl} will not be detected by the tool
because \lstinline{&A[0]} and \lstinline{&A[128]} only have control
dependence on \lstinline{secret}. On the other hand, the tool will
detect the leakage in Fig.~\ref{lst:dat} because \lstinline{secret} is
involved in the computation of \lstinline{&A[i]}.
Furthermore, \lstinline{addr} will be tagged as a \textit{leak point}.
A leak point is a memory location where a leaked value resides.

\begin{figure}[tbp]
  \centering
  \begin{subfigure}{0.4\linewidth}
    \centering
\begin{mycode-s}[
  ]
secret = *addr;
if (secret 
  a = A[0];
else
  a = A[128];
\end{mycode-s}
    \caption{Control flow}
    \label{lst:ctl}
  \end{subfigure}%
  \qquad
  \begin{subfigure}{0.4\linewidth}
    \centering
\begin{mycode-s}[
  ]
secret = *addr;
i += 64 * secret;
a = A[i];
\end{mycode-s}
    \caption{Data flow}
    \label{lst:dat}
  \end{subfigure}
  \caption{Code that leaks \lstinline{secret}.}
\end{figure}


Clueless uses an algorithm based on dynamic information flow tracking
(DIFT) \cite{10.1145/359636.359712,suh2004secure,clause2007dytan}.
DIFT has been successfully applied to prevent attacks on software
\cite{clause2007dytan,lam2006general,suh2004secure,haldar2005dynamic,kong2006improving,qin2006lift}
and has been seen in hardware protection proposals against speculative
execution attacks \cite{stt:MICRO2019}. DIFT tracks information flow
by associating taints with data and propagating the taints according
to the data flow. In addition, Clueless's algorithm needs to
automatically assign taints to data and maintain the taints.

\subsection{Taint assignment}

A new taint is assigned to a memory location whenever a \emph{value}
(i.e., data that should not be used to address memory) is loaded from
that memory location. Each taint is associated to the address of a
value. In the example in Fig.~\ref{lst:addrxy}, suppose that values
reside at memory location \lstinline{addrX} and \lstinline{addrY}, a
new taint \(t_{x}\) is assigned to \lstinline{addrX} when the load
instruction on line 3 executes, and then another taint \(t_{y}\) is
assigned to \lstinline{addrY} when line 4 executes.

Clueless needs to know if the loaded data is a value. Most
contemporary Instruction Set Architectures (ISAs) do not make a
distinction between \emph{value} and \emph{address} loads 
nor between \emph{value} and \emph{address} registers.
As a binary instrumentation tool,
Clueless is clueless about which loads actually load values. Clueless
provides two solutions to this problem.

\begin{mycode}[
  caption={Instructions where \lstinline{addrX} and \lstinline{addrY}
are leak points.},
  label={lst:addrxy},
  float=tbp]
r1 = addrX
r2 = addrY
load rX <- (r1)
load rY <- (r2)
r3 = rX * 64
r4 = r3 + rY
load r5 <- (r4)
\end{mycode}

\paragraph{Everything is a value} One solution is to regard all data
in the memory initially as values, i.e., Clueless assumes nothing in
the memory should be used as a memory address. For every load
instruction, a new taint is assigned to the memory address of the
load. Consequently, all memory locations which contain memory
addresses will be considered as leak points. Clueless effectively
provides a way to classify any data in memory into memory addresses or
non-addresses based on the past execution. This provides a new
perspective to analyse programs: how much of a process's memory is
potentially observable by another process through a cache side
channel? We name this model as \emph{aggregation mode}. Aggregation
mode indicate how visible a program memory can be ---
Section~\ref{sec:aggregating-mode} presents its results.

\paragraph{Users set watchpoints} Another solution is to let users
mark out memory regions that contain values. A new taint is assigned
to the memory address of a load only if that address is within a
marked memory region. Clueless supports this solution by providing an
API that can dynamically register and unregister memory regions to
watch. This requires users to modify the source code of instrumented
programs by inserting Clueless watchpoint API calls. We name this
model as \emph{tracking mode}. Section~\ref{sec:tracking-mode}
presents its results.

\subsection{Taint propagation}
\label{sec:taint-propagation}

Clueless uses bit arrays to store taint sets, where each bit
represents a different taint. With this representation, set union
operations are equivalent to bit-wise or operations, which are
efficient to perform. Each bit array is associated to a register or a
memory location. The number of bits in a bit array is finite and can
be configured when compiling the tool. Consequently, the maximal
number of taints is equal to the number of bits in a bit array.

At instruction level, data-flow can be divided into two categories:
register-register flow and register-memory flow. One of the main
differences of the two categories in the context of dependence
tracking is that the space required by register-register flow tracking
is upper-bounded by the number of architectural registers while that
of register-memory flow tracking is upper-bounded by the number of
virtual memory locations. For example, pairs of a load and a store
(both cause register-memory flow) can copy some data throughout the
entire virtual memory and result in every memory location being
tainted by the taints of the data, requiring enormous amount of space
to store the taint sets. This might not be an issue when a few pieces
of data are tracked because the data are not likely to flow through a
large part of the memory. When the numbers of tracking points are
large, however, the space overhead makes complete tracking of
register-memory flow impractical. This is the case for Clueless in
aggregating mode --- it regards everything in the memory as a value and
tracks the entire memory. On the other hand, storing a taint set for
each register requires much less space because the number of
architectural registers is low.

\begin{table*}[tbp]
  \centering
  \caption{Example Taint Propagation}
  \begin{tabular}{|l|cccc|l|}
    \hline
    Instruction & \texttt{rX} & \texttt{rY} & \texttt{r3} & \texttt{r4} & Remark \\
    \hline
    \texttt{r1 = addrX} & \(\{\}\) & \(\{\}\) & \(\{\}\) & \(\{\}\) & \\
    \texttt{r2 = addrY} & \(\{\}\) & \(\{\}\) & \(\{\}\) & \(\{\}\) & \\
    \texttt{load rX <- (r1)} & \(\{t_{x}\}\) & \(\{\}\) & \(\{\}\) & \(\{\}\) & \(t_{x}\) associated to \texttt{addrX} \\
    \texttt{load rY <- (r2)} & \(\{t_{x}\}\) & \(\{t_{y}\}\) & \(\{\}\) & \(\{\}\) & \(t_{y}\) associated to \texttt{addrY} \\
    \texttt{r3 = rX * 64} & \(\{t_{x}\}\) & \(\{t_{y}\}\) & \(\{t_{x}\}\) & \(\{\}\) & \\
    \texttt{r4 = r3 + rY} & \(\{t_{x}\}\) & \(\{t_{y}\}\) & \(\{t_{x}\}\) & \(\{t_{x},t_{y}\}\) & \\
    \texttt{load r5 <- (r4)} & \(\{\}\) & \(\{\}\) & \(\{\}\) & \(\{\}\) & \texttt{addrX}, \texttt{addrY} tagged \\
    \hline
  \end{tabular}
  \label{tab:propagation}
\end{table*}

\paragraph{Tracking dependences via registers} Clueless tracks
register-register flow by examining instructions, identifying source,
destination, and memory addressing registers and following propagation
rules. Table~\ref{tab:propagation} demonstrates how taints propagate
through the registers as instructions from Fig.~\ref{lst:addrxy} are
executed. The propagation rules are listed below:
\begin{enumerate}
\item For each load instruction, the taint set of its destination
  registers becomes either a singleton or an empty set. If a value is
  loaded, the taint set is a singleton whose element is the new taint
  associated to the value's address. If what is loaded is not a value,
  the taint set is the empty set. \newcounter{ppload}
  \setcounter{ppload}{\theenumi}
\item For instructions that set their destination registers to a
  constant (e.g. \lstinline{xor} with two same source registers,
  \lstinline{mov} a constant to a register), the taint sets of their
  destination registers become the empty set. \newcounter{ppconst}
  \setcounter{ppconst}{\theenumi}
\item For instructions whose source and destination operands are all
  registers except the instructions in rule \theppconst{}, the taint
  sets of their destination registers become the union of the taint
  sets of their source registers.
\item For load and store instructions, memory addressing registers have
  their taint sets emptied. All the memory addresses associated with
  the emptied taints are tagged as leak points.
  \newcounter{ppmark} \setcounter{ppmark}{\theenumi}
\item For store instructions, if the taint sets of all the memory
  addressing registers are the empty set, the address is no longer a
  leak point and is untagged.

  \newcounter{ppenumi}
  \setcounter{ppenumi}{\theenumi}
\end{enumerate}

\paragraph{Expanding dependence tracking to memory} Using
register-register flow tracking alone, the taint sets of data could be
lost because programs often store some data to the memory, use the
register containing the data for something else, and later reload the
data from the memory. These cases require tracking register-memory
flow to store and reload the taint sets. Two additional propagation
rules are introduced to expand dependence tracking to memory:

\begin{enumerate}
  \setcounter{enumi}{\theppenumi}
\item For each store instruction, the taint set of the memory address
  becomes the taint set of the source register that contains the
  stored data.
\item For each load instruction, in addition to rule \theppload{}, the
  taint set of a destination register becomes the union of the
  resulting taint set from rule \theppload{} and the taint set of the memory
  address.
\end{enumerate}

Although tracking all the register-memory flow using a complete method
is impractical due to the space requirements, it is still important to
track these flows because temporarily storing data to memory is very
common. For this reason, a set-associative cache is used as a best-effort
approach to store the taint sets that are associated with memory
addresses. The cache uses a first-in-first-out replacement policy.
The number of sets as well as the associativity of the cache can be
configured when compiling the tool.

\subsection{Taint maintenance}
Clueless has finite number of taints because of the use of statically
sized bit arrays as taint sets. Therefore, taints must be maintained
and reused. A taint can only be reused when it is in none of the taint
sets. Propagation rule \theppload{}, \theppconst{} and \theppmark{}
are the rules that can empty taint sets and make taints reusable.
Since the addresses associated with the emptied taints are already
tagged as leak points according to propagation rule \theppmark{}, the
emptied taints no longer have useful information, thus can be removed
from all the taint sets, resulting in them immediately becoming
reusable.

Taints can still be exhausted in spite of the recycling. For example,
a program can have a loop that loads many values from the memory and
sums them. In these cases, Clueless makes the taint assigned by the
earliest load available by removing it from all the taint sets.

\subsection{Limitations}
\label{sec:limitations}

\paragraph{No tracking on speculative execution}
Clueless is a binary instrumentation tool. It is not a hardware
simulator and does not obtain micro architectural information such as
instructions executed in speculation. As a result, Clueless cannot
track speculative execution.

\paragraph{Incomplete tracking}
Clueless is a characterisation tool as oppose to a verification tool.
The tracking of Clueless is incomplete. Clueless can track data
dependence within a limited window. The incompleteness is the
consequence of our implementation that uses a finite number of taints
and a finite sized cache. When compiling the tool, users can adjust
these parameters to find the desired size of the tracking window.


\paragraph{Dependencies}
Clueless depends on Intel Pin \cite{intel}. Clueless is compiled into
a shared library and needs to be loaded by Intel Pin. The propagation
algorithms of Clueless is implemented in a platform-independent way,
but Intel Pin only supports instrumentation of IA-32, x86-64 and MIC
ISAs. As a result, Clueless currently only supports these ISAs.

\subsection{Source code}
The source code of Clueless is published under the GNU General Public
License, Version 3. Its git repository is accessible at
\url{https://github.com/xiaoyuechen/dift-addr.git}.

\section{Aggregating mode}
\label{sec:aggregating-mode}


Clueless in aggregating mode regards everything in the instrumented
program's memory as values. Clueless in this mode tags any memory
locations whose data transform into addresses as leak points. In
addition, Clueless collects a set of all memory addresses used by the
program, i.e., addresses used in any memory accessing instructions.
With the set of leak points and the set of all memory addresses, we
could introduce a metric that describes the proportion of data that
are used as addresses for a given execution of a program.

\subsection{The \(\Lambda\) metric}

Let \(L_{i}\) be the set of leak points and \(A_{i}\) be the set of
all addresses after the execution of the \(i\):th instruction of a
program (Trivially, \(L_{i} \subseteq A_{i}\)). Let \(n\) be the number of
instructions of the entire execution of the program, metric
\(\Lambda\) defined by
\[
  \Lambda = \frac{\sum_{i=1}^{n}|L_{i}|}{\sum_{i=1}^{n}|A_{i}|}
\]
indicates the average proportion of data that transform into addresses
during the entire executing of the program. Figuratively, \(\Lambda\) is the
area under the \(|L_{i}|\) curve divided by the area under the
\(|A_{i}|\) curve in Fig.~\ref{fig:al}.

\subsection{\(\Lambda\) of SPEC benchmarks}

We used Clueless's aggregating mode on SPEC 2006 to characterise data
transformation into memory addresses by analysing how \(|A_{i}|\) and
\(|L_{i}|\) change and comparing the \(\Lambda\) values of different
benchmarks. Since Clueless uses incomplete methods to track \(L_{i}\)
while the tracking of \(A_{i}\) is complete, the reported values of
\(\Lambda\) are lower-bounds of the actual \(\Lambda\).

\begin{figure}[tbp]
  \includegraphics[width=0.95\columnwidth]{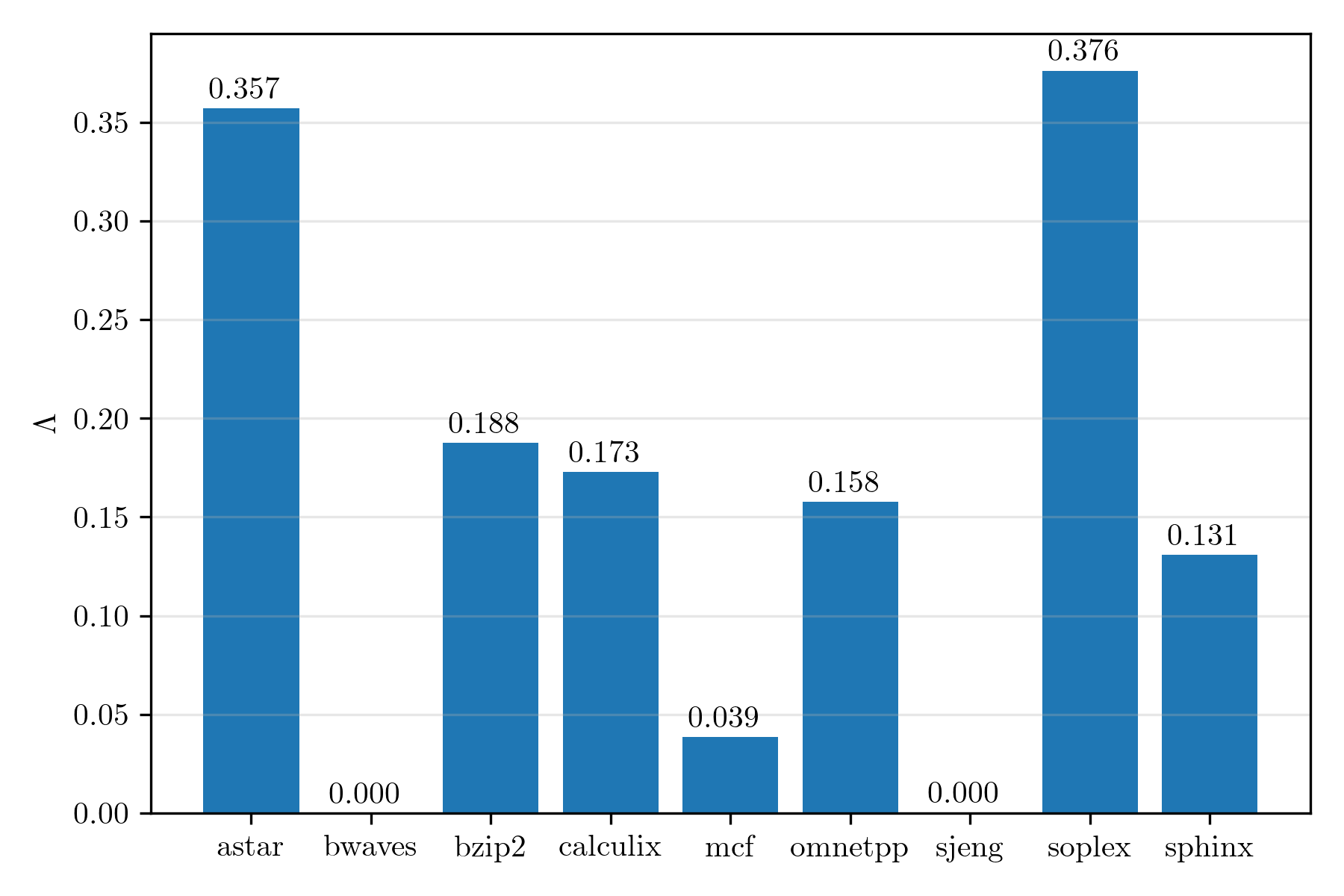}
  \caption{\(\Lambda\) of SPEC benchmarks.}
\label{fig:lambda}
\end{figure}

The prevalence of data-address transformations, indicated by
\(\Lambda\), is an innate property of a program. Fig.~\ref{fig:lambda} shows
the values of \(\Lambda\) of different benchmarks programs. The astar
program and soplex program use more than one third of their memory to
store addresses. In the bwaves program and sjeng program, on the other
hand, such transformations are rarely seen.

\subsection{A closer look}

For more insights into data-address transformation, we further study
how much data are transformed into addresses at each point of
execution of some benchmark programs.

\begin{figure*}[tbp]
  \begin{subfigure}{0.23\textwidth}
    \includegraphics[width=\columnwidth]{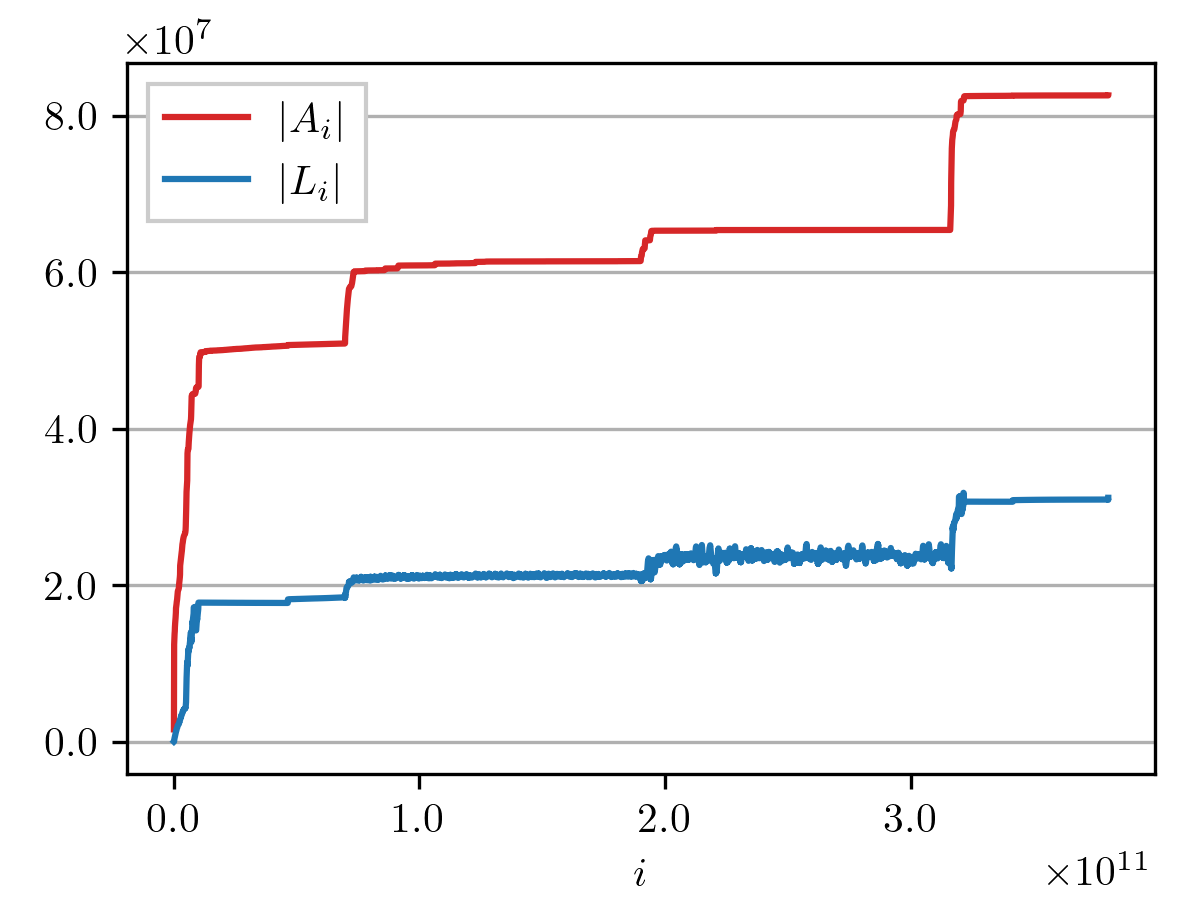}
    \caption{astar}
    \label{fig:astar}
  \end{subfigure}
  \begin{subfigure}{0.23\textwidth}
    \includegraphics[width=0.95\columnwidth]{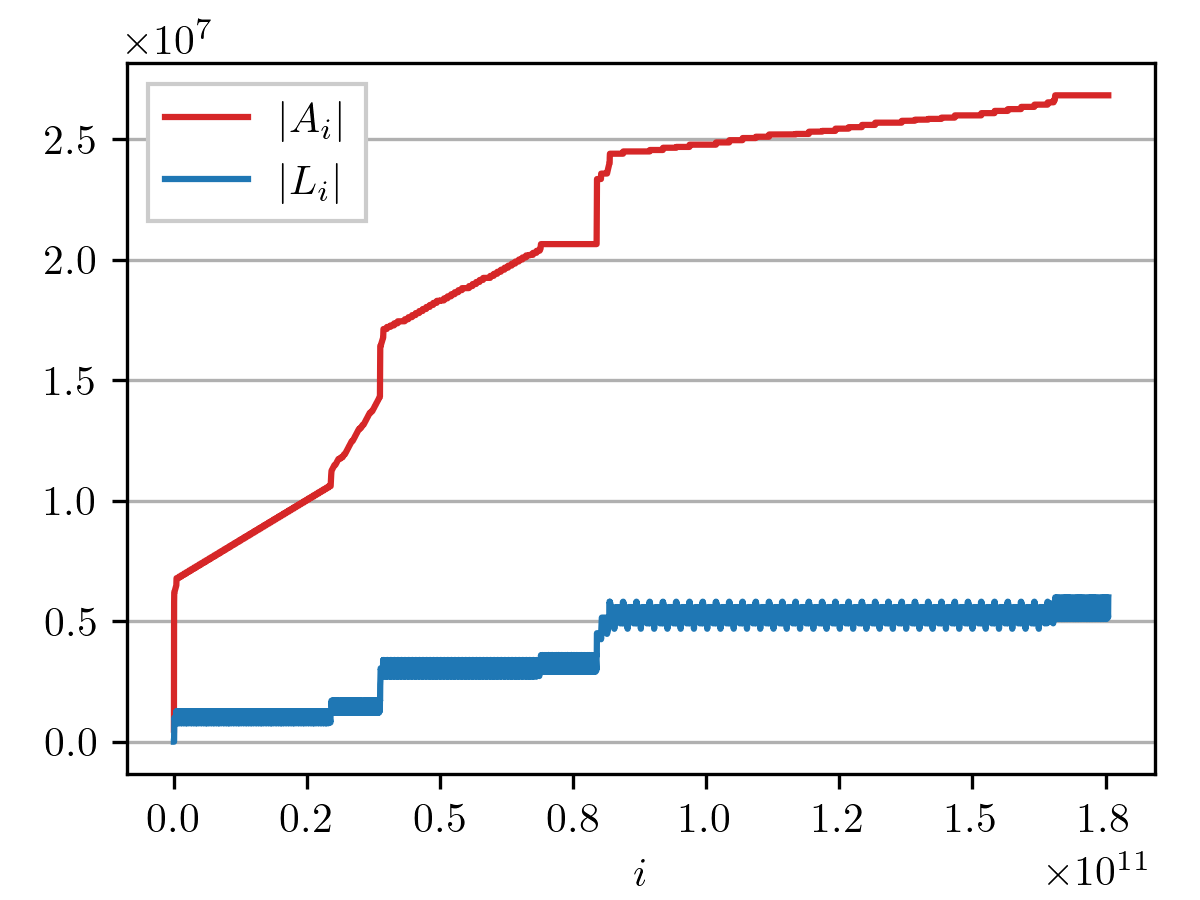}
    \caption{bzip2}
    \label{fig:bzip2}
  \end{subfigure}
  \begin{subfigure}{0.23\textwidth}
    \includegraphics[width=0.95\columnwidth]{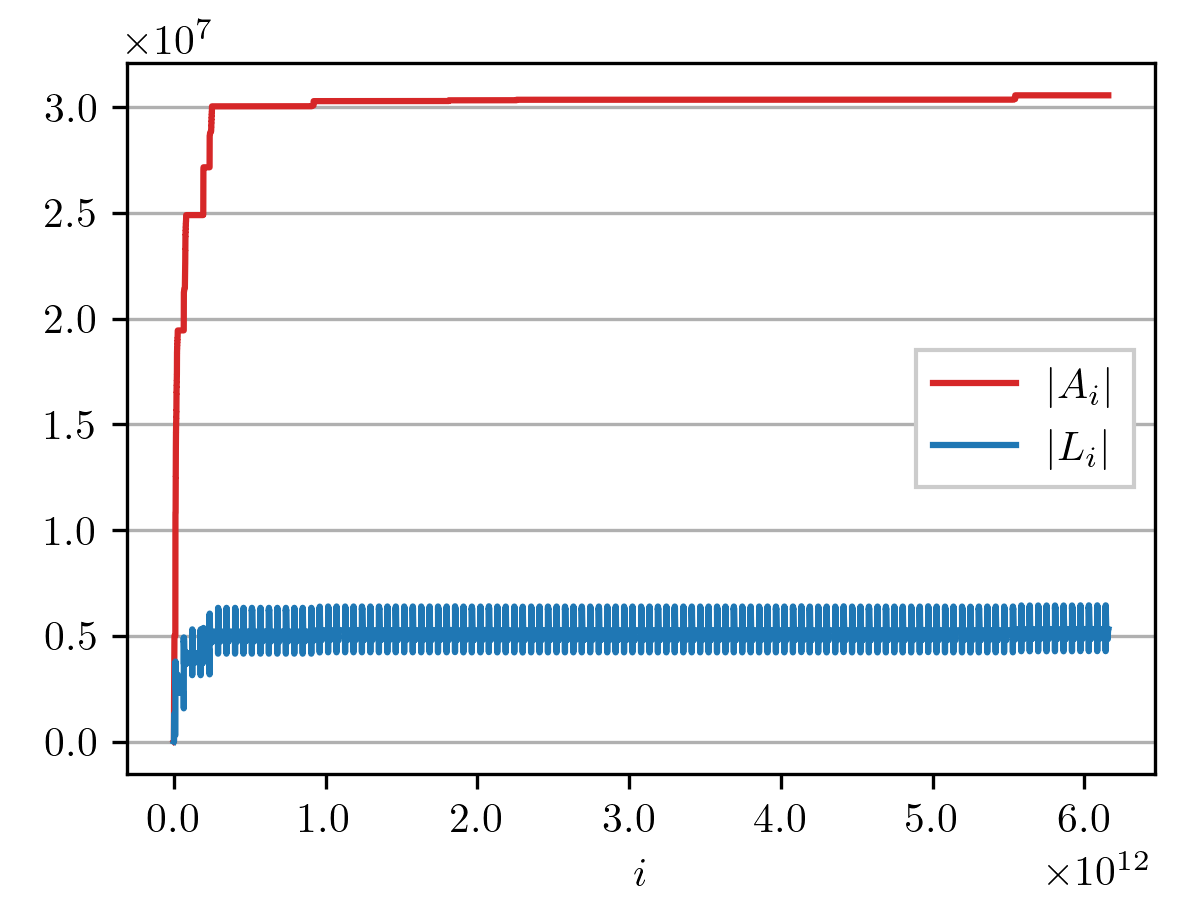}
    \caption{calculix}
    \label{fig:calculix}
  \end{subfigure}
  \begin{subfigure}{0.23\textwidth}
    \includegraphics[width=0.95\columnwidth]{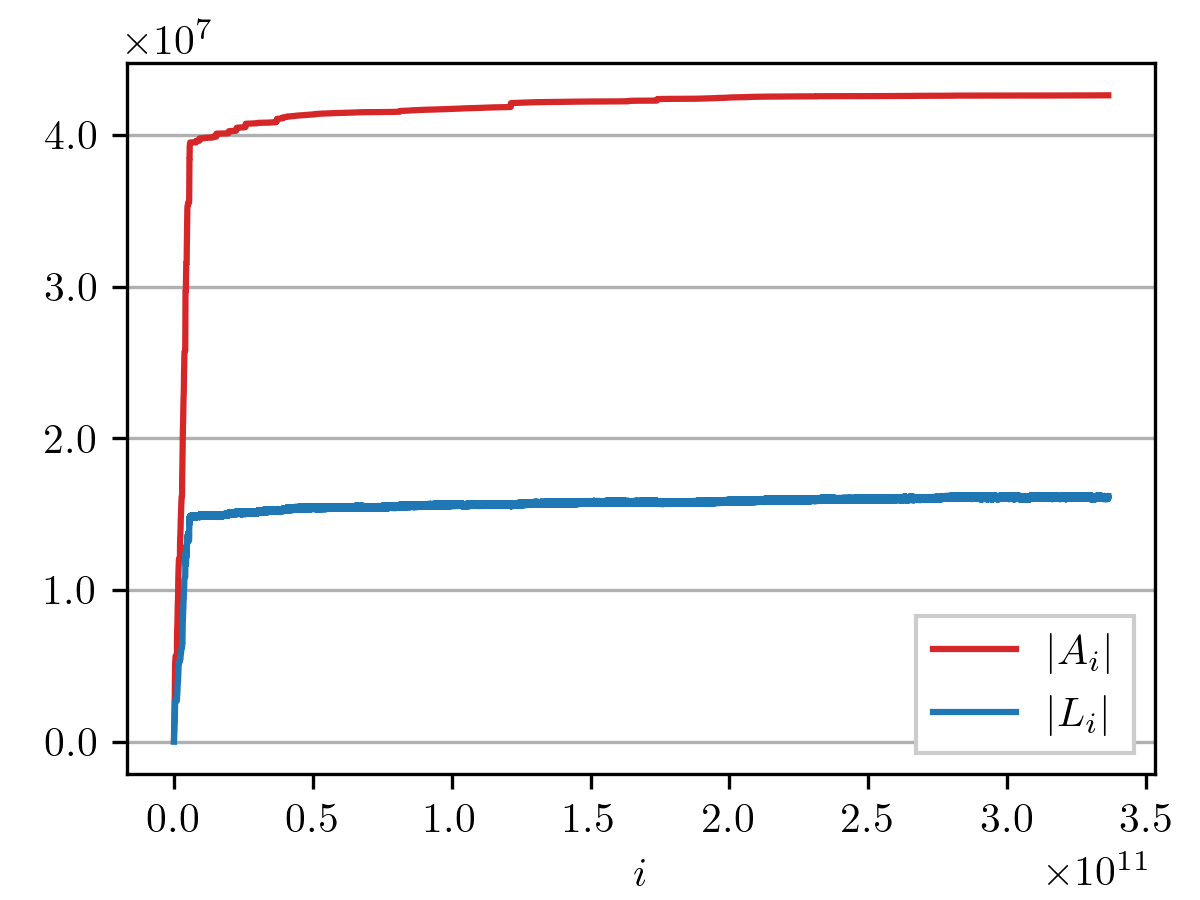}
    \caption{soplex}
    \label{fig:soplex}
  \end{subfigure}
  \caption{\(|A_{i}|\) and \(|L_{i}|\) of benchmark programs.}
  \label{fig:al}
\end{figure*}

\paragraph{Astar} Fig.~\ref{fig:astar} shows how \(|A_{i}|\) and
\(|L_{i}|\) change during an execution of the astar program. Note that
\(|A_{i}|\) increases monotonically while \(|L_{i}|\) does not. When
an address \(l \in L_{i}\) is written to by the \((i+1)\):th instruction,
\(L_{i+1} = L_{i} \setminus \{l\}\). This is very common when \(l\) is a stack
address, as the stack memory tends to be rewritten often. In the astar
program, \(|L_{i}|\) fluctuates when
\(i \in [2\times10^{11},~3\times10^{11}]\). The cause of such fluctuations is that
the same blocks of memory containing addresses are repeatedly loaded
from and written to.

\paragraph{Bzip2} Fig.~\ref{fig:bzip2} shows that the bzip2 program
periodically store new addresses to the same blocks of memory. One
common data-address transformation pattern can be found when
\(i \in [0,~3.9 \times 10^{10}]\),
\(i \in [3.9 \times 10^{10},~7.9 \times 10^{10}]\), and
\(i \in [7.9 \times 10^{10},~1.75 \times 10^{11}]\) --- a rapid increase in
\(|L_{i}|\) which then fluctuates periodically, followed by another
rapid but smaller increase in \(|L_{i}|\) and fluctuates periodically
again. The cause for the repeated pattern could be that the bzip2
program reallocates memory to store memory addresses, but the same
algorithm is used on the reallocated memory.

\paragraph{Calculix} Fig.~\ref{fig:calculix} shows that the calculix
program has an obvious periodic memory access pattern. After the
initial increase of both \(|A_{i}|\) and \(|L_{i}|\), \(|A_{i}|\)
becomes stable while \(|L_{i}|\) becomes periodic. The amplitude of
\(|L_{i}|\) is relatively large at approximately
\(2.1 \times 10^{6}\), indicating that blocks containing
\(2.1 \times 10^{6}\) addresses are repeatedly written with new addresses.

\paragraph{Soplex} Fig.~\ref{fig:soplex} shows how \(|A_{i}|\) and
\(|L_{i}|\) of the soplex program change. After the initial increase,
both \(|A_{i}|\) and \(|L_{i}|\) become stable. This does \emph{not}
mean that data in this program are transformed to memory addresses
only once. After the data are tagged, they may still be transformed
into memory addresses multiple times in different ways, but \(L_{i}\)
would remain the same. The stable \(|L_{i}|\) only indicates that no
new data are tagged, and no tagged memory location is written to.

\section{Tracking mode}
\label{sec:tracking-mode}

Clueless in tracking mode allows users to dynamically register and
unregister watchpoints, i.e., memory blocks that contain values. If
data from any watchpoint are transformed into memory addresses,
Clueless will provide a detailed diagnose on each leakage. The
diagnostic information include the leak point, the memory address that
the value in the leak point transforms into, a trace of instructions
that shows the value-address transformation, and the relevant routine
and image names where the leakage happens. This mode could be used to
test the side-channel vulnerability of programs and help understand
where and how secrets are leaked if such vulnerability exists.

\subsection{The micro benchmark}

We demonstrate Clueless in tracking mode on a micro benchmark program
in Fig.~\ref{lst:micro}. Array \lstinline{T[]} and
function \lstinline{foo} are defined in a shared library the victim
program links against. The victim program has a secret stored in
the \lstinline{s[]} array. The victim calls \lstinline{foo} with the
secret as its parameter. Function \lstinline{foo} loads each byte of
its parameter array, multiplies the byte value by 64, and uses the
result as the index of a constant array \lstinline{T[]} to do some
lookup.

This program is vulnerable to side channel attacks such as Flush +
Reload \cite{yarom2014flush+}. The attacking program may
\lstinline{mmap} the shared library and flush the cache lines
containing \lstinline{T[]}, wait for the victim to call the
\lstinline{foo} function, and measure the time to reload the cache
lines to find out which lines are accessed by \lstinline{foo}.
Assuming that the victim's machine has 64-byte cache lines, the
attacker can recover the secret completely --- each access of
\lstinline{T[s[i]*64]} will be on a different cache line, so each byte
of the secret can be computed using \lstinline{(l-T)/64} where
\lstinline{l} is the address of an accessed line. The offset of
\lstinline{T} can also be found trivially because it is just a symbol
in a shared library. In our example, \lstinline{T} has an offset of
\lstinline{0x2020}.

\begin{mycode}[
  caption={The micro benchmark program.},
  label={lst:micro},
  float=tbp
  ]
// The shared library:
const uint8 T[256 * 64];
int foo (const uint8 *s, size_t n) {
  for (size_t i = 0; i < n; ++i) {
      uint8 r = T[s[i] * 64];
      ...
    }
}

// The victim program:
uint8 s[] = "cLUe";
foo (s, sizeof (s) - 1);
\end{mycode}

\subsection{Pinpointing the leakage}

\begin{figure}[tbp]
  \includegraphics[width=0.95\columnwidth]{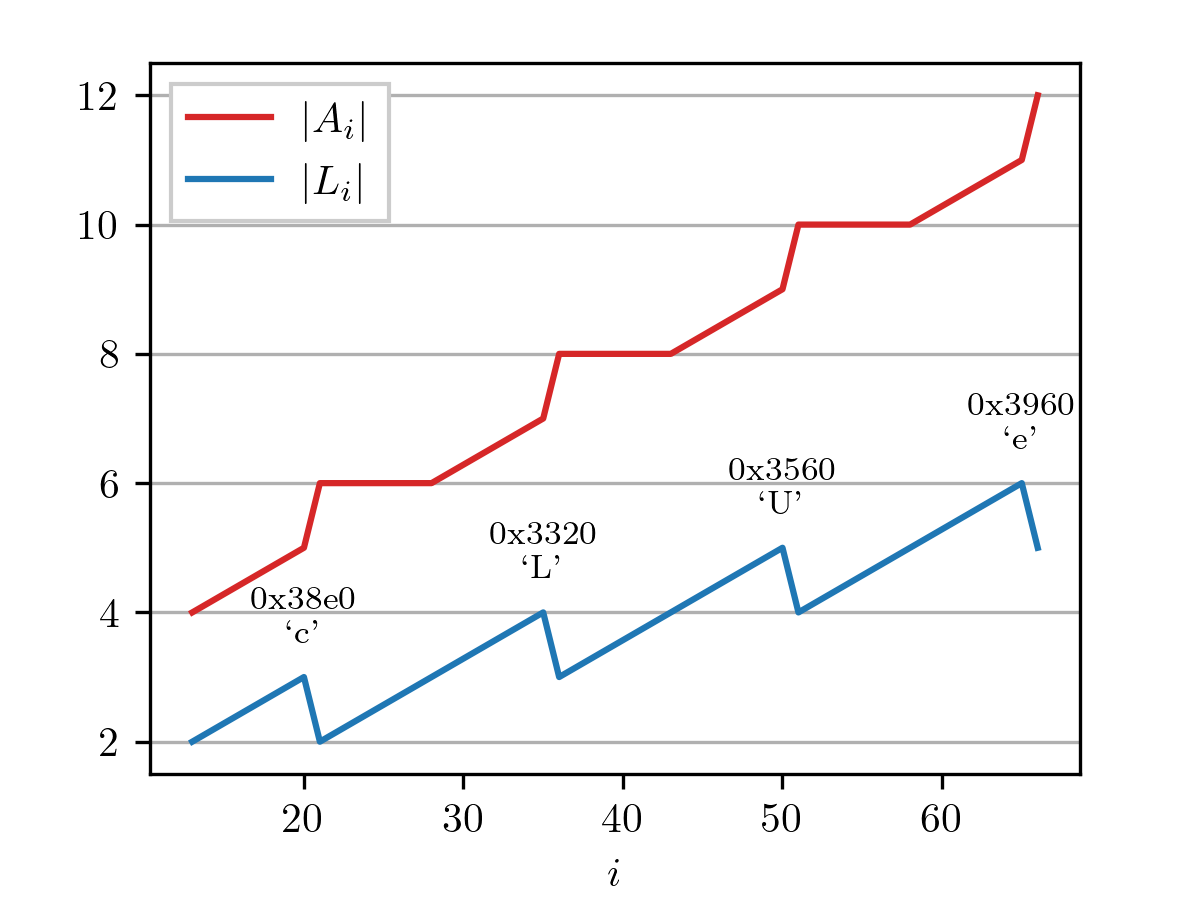}
  \caption{\(|A_{i}|\) and \(|L_{i}|\) of the micro benchmark program.}
\label{fig:micro}
\end{figure}

Clueless's aggregation mode can be used to characterise the micro
benchmark program. Fig.~\ref{fig:micro} shows how its \(|A_{i}|\) and
\(|L_{i}|\) change. With Clueless in tracking mode, we can pinpoint
which increase of \(|L_{i}|\) results in the tracked secrets being
leaked. The annotations in Fig.~\ref{fig:micro} reveal the content of
the leaked secrets at the point of their leakage. For
example, \lstinline{0x38e0} is leaked when \(i=20\),
i.e., \lstinline{&T[s[0]*64]} evaluates to \lstinline{0x38e0} (after
subtracting the image load offset), and is used as a memory address in
a load. To recover \lstinline{s[0]}, we
compute \lstinline{(0x38e0-0x2020)/64} and yield \lstinline{0x63},
which is the ASCII code for `c'.

\subsection{Tracing the transformation}

\begin{figure}[btp]
  \centering
  \begin{subfigure}{0.71\columnwidth}
\begin{mycode-s} [frame=tbl]
0x7fff41801683 { 0 } -> 
[ 
\end{mycode-s}
    \caption{Propagation}
    \label{lst:micro-trace-pp}
  \end{subfigure}%
  \begin{subfigure}{0.34\columnwidth}
\begin{mycode-s} [
  numbers=none,
  xleftmargin=0pt
]
movzbl (
movsbl 
shl    $0x6,
cltq
add    
mov    
mov    (
\end{mycode-s}
    \caption{Instruction}
    \label{lst:micro-trace-ins}
  \end{subfigure}
  \caption{Tracing the micro benchmark.}
  \label{lst:micro-trace}
\end{figure}

For further understanding on the leakage, Clueless gives a trace of
propagation that causes it. Fig.~\ref{lst:micro-trace-pp} shows the
part the propagation trace that causes \lstinline{s[0]} to leak, and
Fig.~\ref{lst:micro-trace-ins} shows the corresponding instructions.

The propagation trace has the following syntax:
\begin{itemize}
\item \lstinline!{ NUM, .. }! is a taint set,
  e.g., \lstinline!%rax { 0, 3 }! means register \lstinline{rax} has
  taint set \(\{t_{0}, t_{3}\}\).
\item \lstinline{->} represents data flow and taint propagation,
  e.g., \lstinline!%rdx { 0 } -> %rcx! means data flow from
  register \lstinline{rdx} to register \lstinline{rcx}, and taint
  \(t_{0}\) propagates to register \lstinline{rcx}.
\item \lstinline![ REG { NUM, .. } ]! represents a register being used
  as a memory address, followed by an equal sign and the effective
  address, e.g., \lstinline![ %rcx { 0 } ] = 0x55baae46d8e0 ->! means
  that register \lstinline{rcx} whose taint set is \({t_{0}}\) is used
  as a memory address in a load.
\end{itemize}

By analysing the trace of the micro benchmark, we find that the
instruction at line 1 loads \lstinline{&s[0]} which
is \lstinline{0x7fff41801683} with taint set \(\{t_{0}\}\) to
register \lstinline{rax}. The following 3 instructions propagate
\(\{t_{0}\}\) from register \lstinline{rax} to itself. Then the
instruction at line 5 propagates \(\{t_{0}\}\) to
register \lstinline{rdx} by adding register \lstinline{rax} to it.
\(\{t_{0}\}\) is further propagated to register \lstinline{rcx} which
is eventually used as the memory address \lstinline{0x55baae46d8e0}.

\section{Case study: AES}

We have seen how the micro benchmark in Section
\ref{sec:tracking-mode} could leak secrets due to its value-address
transformations. Some implementations of Advanced Encryption Standard
(AES) \cite{aes2001} are susceptible to cache side-channel attacks for
the same reason. These implementations often depend on large tables to
speed up the encryption process \cite{hamburg2009accelerating}. If
encryption keys are transformed into indices of large tables for
lookups, attackers may partially or completely recover the keys by
observing the corresponding cache state changes. Numerous attacks on
AES exploiting this class of vulnerability have been demonstrated in
the past
\cite{aciiccmez2007cache,gullasch2011cache,osvik2006cache,bonneau2006cache}.
Different implementations of AES have also been proposed to protect
against such attacks while retaining or improving the speed of
encryption
\cite{osvik2006cache,hamburg2009accelerating,gueron2010intel}.

In this case study, we use Clueless in tracking mode to analyse three
different implementations of AES present in OpenSSL 3.0.3 --- T-table,
Vector Permutation AES (VPAES), and Intel Advanced Encryption Standard
New Instructions (AES-NI). The expanded encryption key is set as the
watchpoint in order to observe if it is transformed into memory
addresses.

\subsection{T-table}
The T-table implementation of AES in OpenSSL uses 9 T-tables, i.e.,
pre-computed lookup tables, with 8 of them being 8kiB each and 1 of
them being 2kiB. The encryption key is first expanded to round keys.
The first round key is combined with the 16-byte plaintext using
\lstinline{xor} to form the initial state vector. The elements in the
state vector are then used as indices of the T-tables to look up
values which are combined with the next round key to form the next
state vector. This implementation could be easily broken using
Prime+Probe, with the 128-bit encryption key fully recovered after
only 300 encryptions \cite{osvik2006cache}.

Clueless detects the potential leak and marks all the bytes of the key
as leak points. In addition, Clueless gives a propagation trace that
shows how the key is transformed into memory addresses.
Fig.~\ref{lst:aes-t-pp} shows the propagation trace of the first round
of the encryption (all rounds are similar) while
Fig.~\ref{lst:aes-t-ins} shows the corresponding instructions. The
traces shows that the first round key is leaked as follows:

\begin{enumerate}
\item Instructions from line 1 to 4 load the first round key
  and \lstinline{xor} them with the plaintext in
  register \lstinline{rax}, \lstinline{rbx}, \lstinline{rcx},
  and \lstinline{rax} to form the initial state vector. Different
  taints are assigned to the watched memory locations and propagated
  to the destination registers.
\item Instructions from line 5 to 8 extracts some bytes from the
  initial state vector and copy them to a different set of registers
  (\lstinline{r10}, \lstinline{r11}, \lstinline{r12}
  and \lstinline{r8}). Taints propagate from the source registers to
  the destination registers.
\item Instructions from line 13 to 16 use the extracted bytes to
  perform lookup in one of the T-tables. The tainted registers are
  used to form effective addresses. At this point, Clueless marks the
  memory locations of the first round key as leak points.
\end{enumerate}

\begin{figure*}[tbp]
  \centering
  \begin{subfigure}{0.35\linewidth}
\begin{mycode-s} [
  frame=tbl,
  ]
0x7ffc4afe86b0 { 1 } -> 
0x7ffc4afe86b4 { 2 } -> 
0x7ffc4afe86b8 { 3 } -> 
0x7ffc4afe86bc { 4 } -> 
[ 
[ 
[ 
[ 
\end{mycode-s}
\caption{Propagation}
\label{lst:aes-t-pp}
\end{subfigure}%
\begin{subfigure}{0.3\linewidth}
  \begin{mycode-s} [
  numbers=none,
  xleftmargin=0mm,
  framexleftmargin=0mm,
  ]
xor    (
xor    0x4(
xor    0x8(
xor    0xc(
movzbl 
movzbl 
movzbl 
movzbl 
movzbl 
movzbl 
shr    $0x10,
movzbl 
movzbl (
movzbl (
movzbl (
movzbl (
\end{mycode-s}
\caption{Instruction}
\label{lst:aes-t-ins}
\end{subfigure}
  \caption{Tracing the T-table implementation.}
  \label{lst:trace}
\end{figure*}

\subsection{VPAES}

VPAES is a technique for accelerating AES using vector permute
instructions. It avoids key-dependent memory references thus being
immune to known cache-timing attacks \cite{hamburg2009accelerating}.
The hardware must support vector permutation instructions to use
VPAES.

Clueless in tracking mode shows that no part of the key is leaked as
memory addresses in OpenSSL's VPAES implementation. The reported
propagation trace in Fig.~\ref{lst:aes-vp-pp} indicates that the key
is loaded (e.g., by instruction at \lstinline{86e1}) and combined in
different rounds, but no part of the key is used to reference memory.

\begin{mycode}[
  caption={Tracing the VPAES implementation.},
  label={lst:aes-vp-pp},
  numbers=none,
  xleftmargin=0mm,
  framexleftmargin=0mm,
  float=tbp,
  % language={[x86masm]Assembler}
]
0x7fff609ec660 { 1 } -> %xmm5
%xmm2 %xmm5 { 1 } -> %xmm2
%xmm0 %xmm2 { 1 } -> %xmm0
%xmm1 %xmm0 { 1 } -> %xmm1
%xmm1 { 1 } -> %xmm1
%xmm0 { 1 } -> %xmm0
%xmm5 %xmm0 { 1 } -> %xmm5
%xmm0 %xmm1 { 1 } -> %xmm0
%xmm3 %xmm1 { 1 } -> %xmm3
%xmm3 %xmm5 { 1 } -> %xmm3
%xmm4 %xmm0 { 1 } -> %xmm4
%xmm4 %xmm5 { 1 } -> %xmm4
%xmm2 %xmm3 { 1 } -> %xmm2
%xmm2 %xmm0 { 1 } -> %xmm2
%xmm3 %xmm4 { 1 } -> %xmm3
...
%xmm3 %xmm1 { 1 2 ... 10 } -> %xmm3
%xmm4 %xmm2 { 1 2 ... 10 } -> %xmm4
%xmm4 %xmm5 { 1 2 ... 11 } -> %xmm4
%xmm0 %xmm3 { 1 2 ... 10 } -> %xmm0
%xmm0 %xmm4 { 1 2 ... 11 } -> %xmm0
%xmm0 { 1 2 ... 11 } -> %xmm0
\end{mycode}

\subsection{AES-NI}

Intel introduced AES-NI instruction sets in 2010 to provide direct
hardware support for AES. These new instructions run in
data-independent time and do not use tables \cite{gueron2010intel}.
AES-NI can encrypt an entire round with a single instruction. AES
implementations that properly use AES-NI should be immune to cache
side-channel attacks as no cache is involved in these instructions.

The AES implementation in OpenSSL that uses AES-NI has not been found
to transform the key into memory addresses. The propagation trace in
Fig.~\ref{lst:aes-ni-pp} shows that no part of the key is used to
reference memory.

\begin{mycode}[
  caption={Tracing the AES-NI implementation.},
  label={lst:aes-ni-pp},
  numbers=none,
  xleftmargin=0mm,
  framexleftmargin=0mm,
  float=tbp,
  % language={[x86masm]Assembler}
]
0x7ffd85426fd0 { 1 } -> %xmm0
0x7ffd85426fe0 { 2 } -> %xmm1
%xmm2 %xmm0 { 1 } -> %xmm2
%xmm2 %xmm1 { 1 2 } -> %xmm2
0x7ffd85426ff0 { 3 } -> %xmm1
%xmm2 %xmm1 { 1 2 3 } -> %xmm2
0x7ffd85427000 { 4 } -> %xmm1
...
%xmm2 %xmm1 { 1 2 3 ... 10 } -> %xmm2
0x7ffd85427070 { 11 } -> %xmm1
%xmm2 %xmm1 { 1 2 ... 11 } -> %xmm2
%xmm0 { 1 } -> %xmm0
%xmm1 { 11 } -> %xmm1
%xmm2 { 1 2 ... 11 } -> 0x55feae1dd130
\end{mycode}

\section{Conclusion and future work}

We have presented Clueless: a tool characterising values leaking as
addresses. Using Clueless in aggregating mode, we have characterised,
for the first time, the amount of data values that transformed into
memory addresses in SPEC 2006 benchmark programs. Some benchmark
programs use more than one third of accessed memory to reference
memory. Clueless in tracking mode has provided the traces of how
secrets propagate and leak in a micro benchmark and AES
implementations in OpenSSL. The T-table implementation of AES exhibits
potential vulnerabilities to cache side-channel attacks while the
VPAES and AES-NI implementations are immune to such attacks.

The ``leaks'' reported by Clueless are to be further studied. We hope
to identify the value-address transformations that would lead to the
danger of leaking sensitive information from the false positives (e.g.
secrets transforming to addresses on the same cache lines). We are
interested in applying similar dynamic information flow tracking
techniques on hardware models to mitigate cache side-channel attacks
such as Specture. The high frequencies of data-address transformations
in some programs also indicate optimisation opportunities in cache
systems: data that would transform to memory addresses may be
associated to the data the transformed addresses point to. This may be
a focus of our future work.

\begin{acks}
This work was supported by Microsoft Research through its EMEA PhD
Scholarship Programme grant no. 2021-020, the Swedish Research Council
(VR) grant 2018-05254, the VINNOVA grant 2021-02422, the SSF grant
FUS21-0067, and Uppsala University funds for Cybersecurity.
\end{acks}

\bibliographystyle{IEEEtran} \bibliography{refs}

\end{document}